# Theory for tailoring sonic devices: Diffraction dominates over Refraction

*N. Garcia[1], M. Nieto-Vesperinas[2], E.V. Ponizovskaya[1] and M. Torres[1*]*

{1} Laboratorio de Fisica de Sistemas Pequenos y Nanotecnologia,
   Consejo Superior de Investigaciones Cientificas ,
   Serrano 114, 28006 Madrid, Spain

{2} Instituto de Ciencia de Materiales de Madrid,
   Consejo Superior de Investigaciones Científicas,
   Campus de Cantoblanco Madrid
   28049, Spain
[*]Permanent address: Instituto de Física Aplicada,
   Consejo Superior de Investigaciones Cientificas ,
   Serrano 114, 28006 Madrid, Spain

*Abstract*

In the moderately long wavelength propagation regime, acoustic crystal devices which size is of several wavelengths are studied by using the finite-difference time domain method. *It is shown by the focusing and imaging of a square shaped lens that the diffractive effects dominate over the refractive ones.* Taking into account the well known Babinet principle, the major role of the device edge diffraction is shown. The first examples of imaging with a sonic plane lens and of an acoustic prism able to change the propagation direction of a plane wave are also presented here.

When the wavelength is comparable with the lattice parameter, new effects of acoustic band gaps [1,2,3], frequency ranges where corresponding wave propagation is forbidden, and sound localization and tunneling [4], appear. But the moderately long wavelength propagation regime (below the first Bragg gap) of periodic media is not trivial and has also been recently utilized to develop appealing and potentially useful photonic [5] and acoustic [6] apparent refractive devices. The refraction effect is based on the fact what wave velocities in the composite and in the surrounding medium are different and an effective refractive index is claimed that can be defined. These refractive properties and effects concern the realm of new photonic crystal optics-like and acoustic crystal sonics-like respectively. Analogously to the traditional optics, the theory of photonic crystal optics is based on the homogenization of periodic composites [5] what requires structures with a very high number of scatterers (several thousands), i.e., devices which size is comparable to several hundreds of wavelengths. However, acoustic periodic composites resembling homogeneous effective media are not realistic due to their huge size, in the case of sound, or to their high internal losses, in the case of the ultrasound propagation in solid composites. On the contrary, the size of realistic sonic devices working in the moderately low frequency region, near the gap, as recently proposed ones [6], is of only several wavelengths. *In such a case, refractive and diffractive effects are intrinsically mixed because the incident wave "sees" and strongly scatters at the device edges.* Furthermore, in the nonlinear neighbourhood of the gap, isofrequency curves are anisotropically star shaped, i.e., the wavevector k is not isotropically conserved [7] although the refraction phenomenon relies on the k conservation. This fact implies the appearance of degenerate difractive indices [7] and constitutes other significative source, or different mechanism, of inner diffraction in the mentioned periodic acoustic composites. Hence, the homogeneous effective medium theory it is not applicable to study the scarcely long wavelength properties of sonic composite devices and an effective refractive index can not be properly defined. On the other hand, the plane wave (PW) method has been used to study the dispersion relations: frequency ω versus the wavenumber *k* in acoustic periodic composites. And, from dispersion relation curves, both phase and group wave velocities are calculated [1,2]. But the PW method calculates the band structure of waves propagating in infinite periodic systems and its results can be only indirectly compared with wave propagation measurements in a finite system. Besides, reflectance calculation are also unavoidable to ensure the necessary sonic transparency of devices [6].

In order to solve these problems, a finite difference time domain (FDTD) method is used here to theoretically study refractive-diffractive sonic composite devices. It allows us to study and design finite devices which size is comparable with only several wavelengths and to simulate the experiments in the same way that they are carried out. Alternately, it is worth to note that the layer multiple scattering method [8] provides also, besides the frequency band structure of an infinite phononic crystal, the transmittance and reflectance of a finite slab of the crystal.

The FDTD method used here has previously been used by us to successfully study elastic band gap materials [3] and elastic wave localization phenomena [9] and is easily applicable to the temporal integration of any acoustic wave propagation equation.

We study the propagation of the sonic wave in a composite consisting of parallel and infinitely long cylinders embedded in a host material. The cylinders with radius *r* are setted as a triangular or a square lattice with lattice constant *a* and form different structures. The sonic wave in the composite is given by

$$\frac{\partial^2 u_i}{\partial t^2} = \frac{1}{\rho}\left(\frac{\partial}{\partial x_i}\left(\lambda \frac{\partial u^l}{\partial x_l}\right) + \left(\frac{\partial}{\partial x_l}\mu\left(\frac{\partial u^i}{\partial x_i} + \frac{\partial u^l}{\partial x_l}\right)\right)\right)$$

where $u_i$ is ith component of the displacement vector u( r), λ ( r) and μ( r) are the Lame coefficients and ρ( r) is the mass density. We assume the propagation in the x-y plane, perpendicular to the cylinders axis, so the wave equation can be split into two independent equations [3]

$$\rho \frac{\partial^2 u_x}{\partial t^2} = \frac{\partial T_{xx}}{\partial x} + \frac{\partial T_{xy}}{\partial y}$$

$$\rho \frac{\partial^2 u_y}{\partial t^2} = \frac{\partial T_{xy}}{\partial x} + \frac{\partial T_{yy}}{\partial y}$$

where

$$T_{xx} = (\lambda + 2\mu)\partial u_x/\partial x + \lambda \partial u_y/\partial y, \quad T_{yy} = (\lambda + 2\mu)\partial u_y/\partial y + \lambda \partial u_x/\partial x$$

$$T_{xy} = \mu(\partial u_x/\partial y + \lambda \partial u_y/\partial x)$$

The longitudinal and the transverse velocities are given by $c_l = \{(\lambda + 2\mu)/\rho\}^{1/2}$ and $c_t = \{\mu/\rho\}^{1/2}$, respectively.

The Eqs.(2,3) was integrated by means of a FDTD scheme [3,9]. It uses the discretization of the equations in both the space and the time domains, and it permits to obtain the real wave pattern scattered on the composite structure. We set the periodic boundary conditions at the boundaries along the direction of the wave propagation and Mur´s first order absorbing boundary conditions [10] were used at the boundaries perpendicular to the wave propagation direction. To find the reflectance we used a Gaussian pulse. The transmission coefficient is found by normalizing the fast Fourier transform of the part of the initial pulse that pass through the structure to the fast Fourier transform of the incident pulse. To obtain the distribution of the wave intensity we use the sinusoidal incident wave with frequency 1700 Hz. The intensity was obtained by integrating the value $u_x^2+u_y^2$ during the wave period. Here we study the case of the aluminum cylinders setted in the air. FDTD method gives a satisfactory convergence if $dt<\{(1/dx)^2+(1/dy)^2\}^{-1}\}^{1/2}a/c_l^{Al}$ [10]. The calculations were done with dx=dy=30 and $dt=5.89 \cdot 10^{-3} a/c_l^{Al}$ that gives a good convergence.

A biconvex cylindrical lens performed with a triangular periodic arrangement of 32 aluminum cylinders in air, similar to that of the experiment [6], has been studied and the corresponding plane wave propagation has been visualized in Fig. 1(a). The frequency of the incident plane wave is 1700 Hz and the filling ratio of our numerical experiment is $\pi/2 \cdot 3^{1/2} (d/a)^2 = 0.37$, where d is the cylinder diameter and a is the lattice parameter. Numbers near isophones lines of Fig. 1(a) (boundaries of regions represented in grey scale) are sound wave intensities in dB. The isophone mapping compares well with the experimental one [6] to obtain experimental incomplete data. From the focal acoustic region to the lowest intensity symmetrical islands, at both sides of the focal region, there is an attenuation of 13.25 dB (Fig. 1(a)) whereas the corresponding experimental value is about 15 dB (Fig. 4 in reference [6]). The focal length f is defined as the point where the maximum intensity appears in Fig.1(a). When the wave impinges from infinity we obtain f=41cm. Nevertheless, our acoustic lens has an aperture D=44cm which is only twice the wavelength, therefore, its impulse response [12] $\lambda\pi \sin(\pi\Delta x/\lambda f)/x$ should have a width $\Delta x=2f(\lambda/D)$. At 1700 Hz, $\lambda=20$cm, this should be $\Delta x=37$cm which agrees perfectly well with the peak width of Fig. 1(a).

By using the same periodic composite topology, we also studied a sonic crystal slab with parallel faces (Fig. 1(b)) and very surprisingly we also have a "lens effect" with a square shaped lens. This finding, absolutely unnoticed in recent experiments [6], can only be explained by diffraction! The long *focal length* of this *plane lens* was probably unseen due to the limited size of the anechoid chamber used in [6] This can be a plausible explanation of so remarkable discrepancy between the present theoretical study and the mentioned experimental report [6].The FDTD method allows us to calculate the reflectance of this finite system (Fig. 1(c)). Our theoretical calculations fit well experimental reflectance measurements [6]. However, as above mentioned, our simulated acoustic mapping of the back slab region (Fig. 1(b)) unambiguously exhibits a clear focal region that is absolutely non-existent in the corresponding experimental one [6]. Due to diffraction there is also a peak in the transmission pattern, occurring at about 65 cm, the attenuation from the higher to the lower acoustic intensity regions is 15.5 dB (Fig.1(b)), even higher than the one generated by the former cylindrical lens, and has a width of about 33 cm which is smaller than the expected width of a cylindrical lens of the same focal distance and width (D=32 cm) that would be $\Delta x=80$cm. This is remarkable, since it seems to suggest that at sizes comparable to the wavelength, as used here, it is possible to focus with a wave crystal due to diffraction effects, that could overcome the refractive effects of a conventional lens. This is a very important result revealed by the calculations.

In order to full and deeply explore this unexpected acoustic *plane lens* phenomenon, we present here three conclusive imaging experiments (Fig. 2). In Fig. 2(a), a sonic point source is placed at the position O and the corresponding clear acoustic image is generated at location I. Now, we repeat this theoretical simulation by placing the acoustic point source at the position O, in front of the sonic *plane lens* (Fig. 2(b)), and, surprisingly, a likewise clear image is formed at the point I, as in the former biconvex lens theoretical calculations. Furthermore we made a third experiment (Fig.2c) with a rectangular piece of aluminum and more surprisingly, the result of the experiment is analogous to those of Fig 2(a) and Fig. 2(b). Obviously, the geometrical laws of refraction are not fulfilled between object and image distances showing the nonrefractive nature of the phenomenon. This unambiguously proves the possibility of an imaging mechanism based on exclusive diffraction effects and constitutes a new result that open a wide scenario of potential applications.

To further demonstrate the diffractive strong and major influence of the device edge we illustrate Babinet´s principle that relates the diffraction field setted up by complementary objects. This principle has been proved in optics following Kirchhoff´s diffraction theory [11] and for elastic waves also [13]. We consider two objects for which the transmission functions are complementary, i.e., for uniform illumination, such as:

$q_1(x,y) + q_2(x,y) = 1$

For opaque and transparent objects, this means that the opaque areas of the first are the transparent parts of the second and viceversa.

The diffraction pattern amplitude of the second object is

$Q_2(u,v) = \delta(u,v) - Q_1(u,v)$

So that intensities $|Q_2(u,v)|^2$, $|Q_1(u,v)|^2$ of both complementary objects are equal except at the origin.

In Fig.3 we show the sound diffracted by complementary aluminum screens, i.e., screens such that the opening in the slit corresponds to the aluminum portion in the slab. Babinet´s principle ensures that the addition of both sound amplitude patterns exactly corresponds to the pattern when no screen is present [11]. As can be easily observed in Fig.3 (up), the sound pattern of the massive aluminum slab is analogous to those of the periodic composite slab of Fig.1(b). This fact clearly indicates that the diffraction of the device edge plays the principal role in the phocusing and imaging previously described phenomenom. Although, inner diffraction due to the anisotropic star like shape of the isofrequency curves near the gap [7] can play an important complementary role in the subtle imaging mechanism of the strange *plane lens* phenomenon.

Direct analytical calculation of both external edge like and inner periodically modulated degenerate diffraction mechanisms seems now a formidable task. However, as it has been shown, the FDTD method allow us to study so elusive simulations and tailoring future useful experiments.

The problem to change the propagation direction of a plane wave is more directly related to the only refractive effects even if there is also a diffraction influence, although this is not the major role as in focusing; as we have seen above. So, to unambiguously visualize a refractive effect based on a sonic crystal we present here the first example of an acoustic prism able to change the propagation direction of a plane wave. The device is shaped by adequately truncating a square periodic arrangement of 42 aluminum cylinders in air. The lattice parameter is 6,35cm and the cylinder radius is 2cm. The refraction by the prism of an incident plane acoustic wave at a frequency of 1700Hz is clearly shown in Fig. 4. The refracted wave is also plane being the deviation angle of about $11.5^0$. The averaged angle of the prism is estimated as of $53^0$. If we apply twice the Snell law at both sides of the prism, we calculate the refractive index value n=1.21. Since the FDTD method allow us to calculate the sound velocity inside the acoustic composite, we can alternately calculate the corresponding refractive index value, and we obtain n=1.25. Newly, the slight discrepancy between both estimated refractive indices can be due to anisotropic inner diffraction effects of the weakly modulated structure.

As a conclusion we can state that periodic composite acoustic devices which size is of scarcely several wavelengths are able to produce acoustic images but corresponding properties and therefore the lens image formation is may be due to diffraction not refraction. However, although intrinsically mixed, both diffractive properties and refractive ones can be exploited to change the propagation direction of acoustic plane waves as has been illustrated here with a prism. This ability can be useful in architectural acoustic and to design ultrasonic solid devices. Furthermore we have simulated and illustrated Babinet´s principle for acoustic waves and it should be valid for any kind of waves since it is a property of the Fourier transform. And finally of special interest will be to study the forming images of acoustic, elastic and optical waves by diffractive *plane lens*. Our simulations indicate that diffraction plane lenses of the size of a few wavelengths are possible and this is important and issue of further research.

This work has been supported by the DGICYT.

### *References*
1. M. M. Sigalas and E. N. Economou, J. Sound Vib. 158, 377 (1992); Solid State Commun. 86, 141 (1993); Europhys. Lett. 36, 241 (1996); M. S. Kushwaha et al., Phys. Rev. Lett. 71, 2022 (1993); Phys. Rev. B 49, 2313 (1994); E. N. Economou and M. M. Sigalas, Phys. Rev. B 48, 13434 (1993)
2. W. M. Robertson and W. F. Rudy III, J. Acoust. Soc. Am. 104, 694 (1998); F. R. Montero de Espinosa, E. Jiménez and M. Torres, Phys. Rev. Lett. 80, 1208 (1998); J. V. Sánchez-Pérez et al., Phys. Rev. Lett. 80, 5325 (1998); J. O. Vasseur et al., Phys. Rev. Lett. 86, 3012 (2001)
3. D. García-Pablos et al., Phys. Rev. Lett. 84, 4349 (2000)
4. M. Torres et al., Phys. Rev. Lett. 82, 3054 (1999); S. Yang et al., Phys. Rev. Lett. 88, 104301-1 (2002)
5. P. Halevi, A. A. Krokhin and J. Arriaga, Phys. Rev. Lett. 82, 719 (1999); Appl. Phys. Lett. 75, 2725 (1999), N.Garcia, E.V.Ponizovskaya and J.Q.Xiao, Appl. Phys. Lett. (to be published 2002)
6. F. Cervera et al., Phys. Rev. Lett. 88, 023902-1 (2002)
7. H. Kosaka et al., Phys Rev. B 62, 1477 (2000)
8. I.E. Psarobas et al., Phys Rev. B 62, 278 (2000); Z. Liu et al., Phys. Rev. B 62, 2446 (2000)
9. M. Kafesaki, M. M. Sigalas and N. García, Phys. Rev. Lett. 85, 4044 (2000)
10. A. Taflove, The Finite-Difference Time-Domain Method (Artech House, Boston, 1998)
11. M. Born and E. Wolf, Principles of Optics (Cambridge Univ. Press, Cambridge, 1999)
12. J.W.Goodman, Introduction to Fourier Optics (McGraw-Hill, New York, 1968)
13. A. F. Gangi and B. B. Mohanty, J. Acoust. Soc. Am. 53, 525 (1973)

### *Figure Captions*

Fig.1.- (a) Intensity pattern of an incident sound plane wave in air, at 1700 Hz, refracted and diffracted by a periodic arrangement lens-like of aluminum rods. An extended focus-like region can be seen that is dominated by diffraction, (b) - Intensity pattern of an incident sound plane wave in air, at 1700 Hz, interplaying with a periodic squared shape slab of aluminum rods. Notice that the effect of this finite slab is exactly the same as that of Fig.1(a). We also have focused! displacing a little further from the slab the maximum intensity. In this case we know that diffraction dominates completely. (c)- Reflectance spectrum of the finite sonic crystal slab is represented here

Fig. 2 - Imaging of a point wave source in a biconvex (a) and rectangular periodic composite lens (b). In (c) we show the same surprising imaging effect in a single rectangular piece of aluminum conclusively showing the imaging effect is due to the diffraction at the edge of the device.

Fig.3.- Illustration of Babinet´s principle showing diffraction by complementary aluminum screens. Sound pattern diffracted by a massive aluminum slab (top) and by a slit opened through the aluminum plate (down). This simulation, as well as the squared shape lens, is a proof of the strong dominating diffractive effects.

Fig.4.- Sound refracted by an acoustic prism-like made with a square periodic arrangement of aluminum rod. Diffraction does not prevent that the direction of the incident sound plane wave, at 1700 Hz, clearly changes according to a deviation angle of about $11.5^0$.

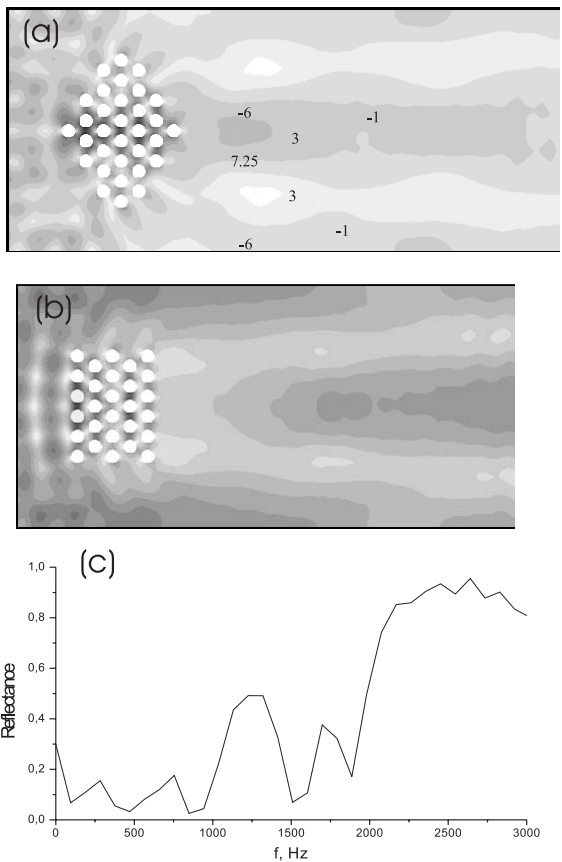
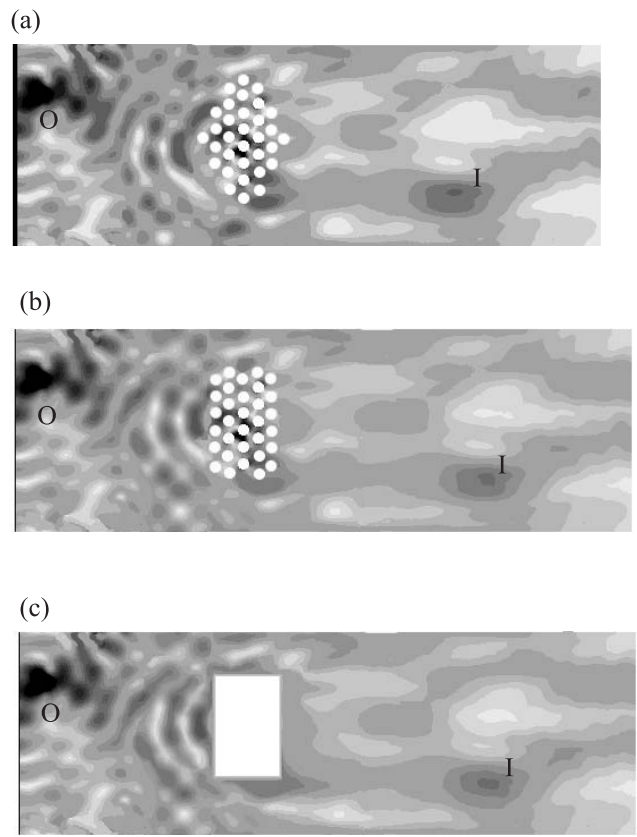

Fig.1            Fig. 2

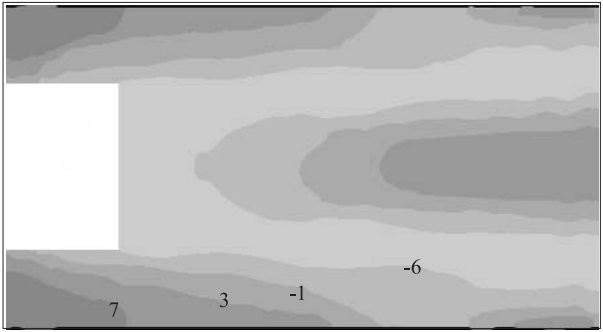

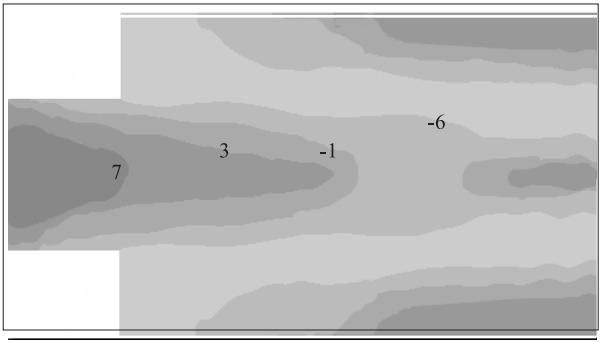

Fig.3

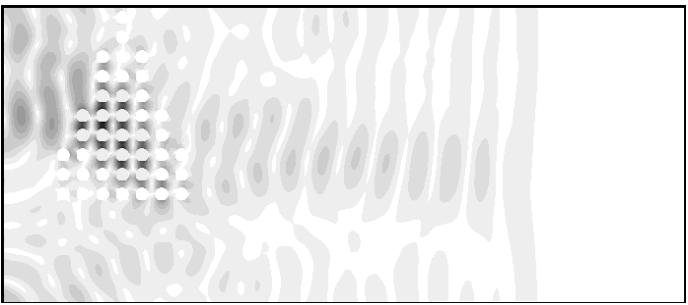

Fig.4